\documentclass{aipproc}
\usepackage{graphicx,setspace}
\usepackage{bm}
\layoutstyle{6x9}
\SetInternalRegister\hbadness{8000}

\newcommand\doingARLO[2][]{%
  \ifx\mmref\undefined #1\else #2\fi
}

\begin{document}

\title{Visco-elastic effects in strongly coupled dusty plasmas}
\author{P. Bandyopadhyay}{
  address={Institute for Plasma Research, Bhat, Gandhinagar - 382428, India},
  }

\iftrue
\author{G. Prasad}{
  address={Institute for Plasma Research, Bhat, Gandhinagar - 382428, India},
  }

\author{A. Sen}{
  address={Institute for Plasma Research, Bhat, Gandhinagar - 382428, India},
  }

\author{P. K. Kaw}{
  address={Institute for Plasma Research, Bhat, Gandhinagar - 382428, India},
  }
\begin{abstract}
We report on experimental evidence of visco-elastic effects in a strongly coupled dusty plasma through investigations of the propagation characteristics of low frequency dust acoustic waves and by excitations of transverse shear waves in a DC discharge Argon plasma.
\end{abstract}
\maketitle
\section{Introduction}
A dusty plasma in the strongly coupled fluid state can develop visco-elastic properties that can significantly influence its collective properties. Recent theoretical studies \cite{rosen2,kaw} have predicted modifications in the dispersive properties of dust acoustic waves (DAWs) as well as the interesting possibility of exciting transverse shear waves (TSWs) in the liquid state. In a series of investigations \cite{pintu1,pintu2} we have tested these ideas and looked for experimental evidence of visco-elasticity in the dusty plasma medium. We highlight two experimental findings in this report : (a) a turnover of the dispersion relation of DAWs caused by strong correlation effects (as distinct from collisional effects) and (b) controlled excitation of TSWs over a range of frequencies and wave numbers. Our experimental findings are in good agreement with theoretical results. \\
\section{Experimental Results and Discussion}
The experiments were carried out in a DC plasma device consisting of a cylindrical chamber filled with Argon gas in which a discharge is struck between a rod shaped anode and the grounded vessel (for more details see \cite{pintu1}) by applying an initial DC voltage of $600$ volts at a gas pressure of $P=1$ mbar. Fine kaolin particles are used as dust and a high density dust cloud is created near the cathode sheath by adjusting the voltage and pressure. The DAW experiments were carried in the parameter range of $V_a=400-550$ volts and $P=0.086-0.5$ mbar. The waves were driven externally by imposing a modulated AC signal of very low frequency (1-3 Hz) on the DC discharge voltage and the wave patterns were recorded on a video and digitally analyzed. The frequency of the applied modulated voltage was varied over a wide range to obtain the dispersive relation. To distinguish between the dispersive effects of dust neutral collisions and dust-dust correlations the dispersion relations were obtained for two different gas pressures~($P=0.086$ and $0.5$ mbar) corresponding to different collisionality regimes. Our results are shown in figure 1(a) and 1(b). The experimentally obtained dispersive characteristics (plotted by filled circle) are compared with the generalized hydrodynamic model results \cite{kaw} (plotted by a solid line).  Both the theoretical and experimental dispersion curves at the two different pressures show a substantial reduction in phase velocity ($V_{ph}=2$ mm/sec), the existence of a regime where $\partial \omega/\partial k < 0$ and a turn over effect at high wave numbers. For comparison we have also plotted the theoretical dispersion relation in which correlation effects are neglected (by a dashed curve)(see fig. 1(a) and 1(b)). It is clear from figure 1(a) that the correlational effects are more dominant at low pressures and  the additional dispersive contribution they provide best explain the dispersion characteristics of DAWs. At high pressure (P=0.5 mbar), the decrease in $\Gamma$ leads to a decrease in the contribution of correlation effects and high neutral pressure makes the collisional contribution to rise as seen from the two theoretical curves of fig 1(b). So in this regime it is difficult to make an unambiguous experimental identification of correlation effects. \par
For the experiments on excitation of transverse shear waves a similar technique of external driving with a modulated voltage has been used. However the TSWs are always found to be excited on top of a spontaneously excited DAW. The TSW wave motion is observed to propagate perpendicularly to the direction of propagation of the acoustic waves with a phase velocity $V_s \sim 7.9$ mm/sec. The linear dispersion characteristics (empty circle), shown in figure 1(c), are compared with the viscoelastic theory (plotted by solid line) proposed in \cite{kaw}. In the high pressure regime, the transverse oscillations get severely damped and are very difficult to sustain. These observations lend experimental support to the existence of visco-elastic effects in a strongly coupled dusty plasma and appear to agree quite well with theoretical predictions.
\begin{figure}
  \resizebox{11pc}{!}{\includegraphics{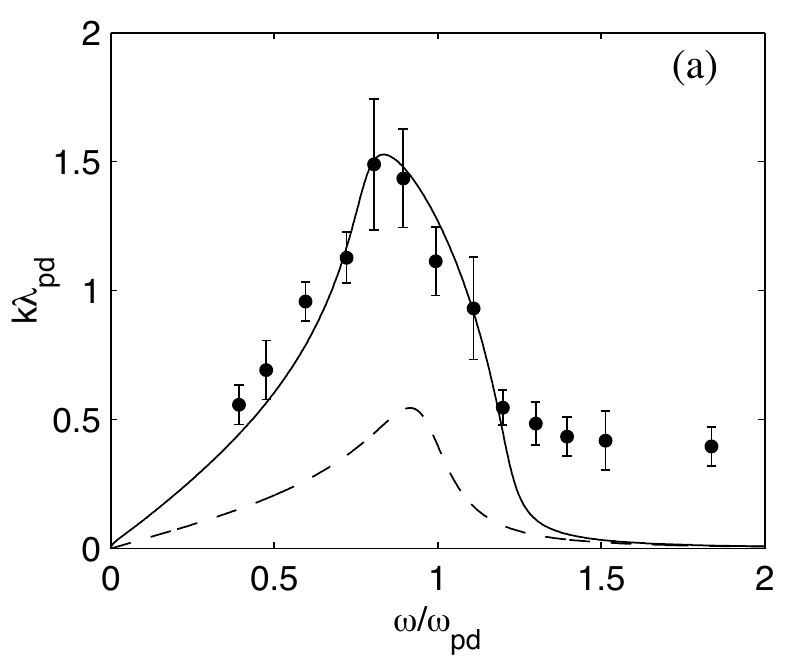}}
  \resizebox{10.9pc}{!}{\includegraphics{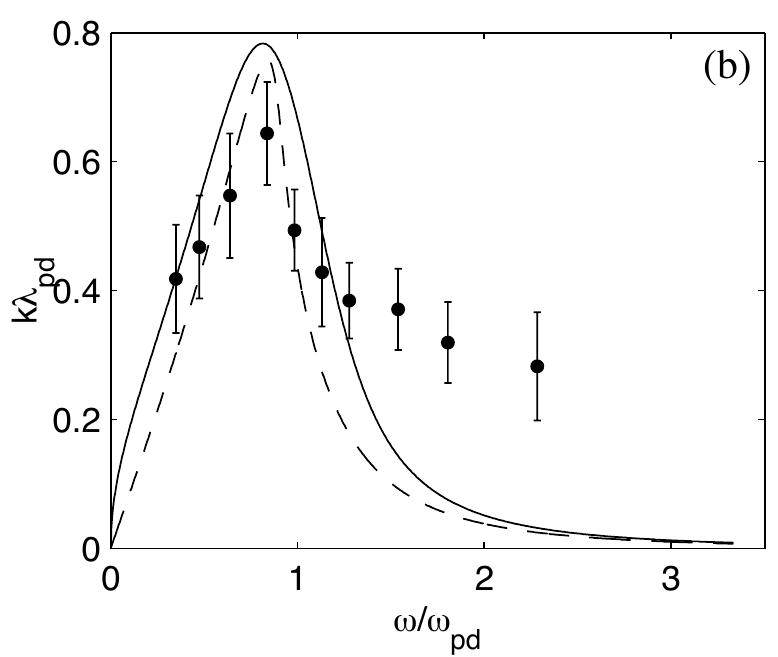}}
  \resizebox{11.7pc}{!}{\includegraphics{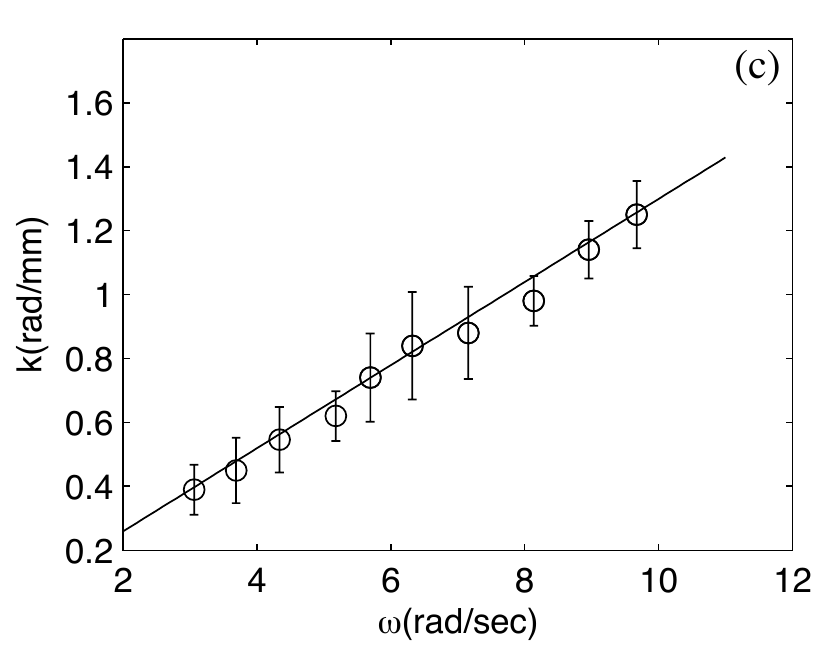}}
  \caption{The theoretical and experimental dispersion characteristics for a) DAWs at P=0.086 mbar b) DAWs at P=0.5 mbar c) Transverse Shear Waves (TSW). }
\end{figure}

\end{document}